\documentclass{INTERSPEECH2023}
\interspeechcameraready

\usepackage{graphicx}
\usepackage{amsmath}
\usepackage{amssymb}
\usepackage{booktabs}
\usepackage{multirow}
\usepackage{footmisc}
\usepackage{algorithm}
\usepackage{algpseudocode}
\usepackage{pifont}
\usepackage{enumitem}

\title{A Training and Inference Strategy Using Noisy and Enhanced Speech as Target for Speech Enhancement without Clean Speech}
\name{Li-Wei Chen$^2$, Yao-Fei Cheng$^2$, Hung-Shin Lee$^1$, Yu Tsao$^2$, and Hsin-Min Wang$^2$}
\address{
$^1$North Co., Ltd., Taiwan \\
$^2$Academia Sinica, Taiwan
}
\email{hungshinlee@gmail.com}
\begin{document}

\maketitle
 
\begin{abstract}
The lack of clean speech is a practical challenge to the development of speech enhancement systems, which means that there is an inevitable mismatch between their training criterion and evaluation metric. In response to this unfavorable situation, we propose a training and inference strategy that additionally uses enhanced speech as a target by improving the previously proposed noisy-target training (NyTT). Because homogeneity between in-domain noise and extraneous noise is the key to the effectiveness of NyTT, we train various student models by remixing 1) the teacher model's estimated speech and noise for enhanced-target training or 2) raw noisy speech and the teacher model's estimated noise for noisy-target training. Experimental results show that our proposed method outperforms several baselines, especially with the teacher/student inference, where predicted clean speech is derived successively through the teacher and final student models.
\end{abstract}

\section{Introduction}
\label{sec:intro}

Speech Enhancement (SE) aims to improve audio quality by removing noise from speech signals. It has a wide range of applications, such as the front-end of automatic speech/speaker recognition systems \cite{Weninger2015,Taherian2020}, where the SE module removes noise from noisy inputs, thereby improving recognition results. The success of current SE development mainly relies on training data containing many pairs of clean and noisy speech \cite{Wang2018,Zhao2018,Choi2019,Zhao2019,Li2020,Koizumi2020,Defossez2020,Hao2021,Pandey2021}. During training, noisy speech is usually synthesized by mixing clean speech and noise so that the SE model can be trained to transform the noisy speech into its corresponding clean speech. This traditional training scheme, called clean-target training (CTT) \cite{Fujimura2021a}, is suitable for various specific applications due to environmental changes. However, due to the higher cost and lower convenience of recording, it is challenging to collect clean speech and in-domain noise in real-world scenarios.

Many methods that operate without clean speech and in-domain noise have recently been proposed to address this problem \cite{Wang2018,Yu2021,Fujimura2021a,Bie2022,Xiang2022,Trinh2022}. Belonging to one branch of \textit{unsupervised} SE\footnote{As described in \cite{Bie2022}, \textit{unsupervised} SE can be defined that the use of paired/parallel noisy and clean speech during training is prohibited or infeasible. Fu \textit{et al.} accordingly further divided \textit{unsupervised} SE tasks into three levels: 1) clean speech or in-domain noise is required; 2) noisy speech is required; and 3) no training data is required \cite{Fu2022}.}, where no subjective/objective speech quality metrics are included as learning reference, and the traditional \textit{ground truth} (i.e., training targets) does not exist in this kind of SE task, researchers have to explore alternatives, which are close to clean speech, for the corresponding noisy speech.

To this end, the noisy-target training (NyTT) method proposed by Fujimura \textit{et al.} \cite{Fujimura2021a} takes a significant step forward by directly treating original noisy speech as the training target. The original noisy speech is used to mix with extraneous noise to form noisier input speech that needs to be enhanced. The extraneous noise can be any corpus of noise recordings other than the in-domain noise. (The authors pretend not to have real in-domain noise for training.) Despite NyTT's competitive performance, it shares a disadvantage with CTT-based SE models trained with paired clean and noisy data. That is, NyTT only performs well when the extraneous noise is close to the realistic in-domain noise contained in the training/test noisy speech. If the extraneous noise is not similar to the in-domain noise, the out-of-domain (OOD) issue can easily distract the processing power of the NyTT model because it has to deal with different noise than the noise seen in training.

To overcome the OOD problem, various unsupervised algorithms have been proposed. For example, mixture invariant training (MixIT) in speech separation enables unsupervised domain adaptation and learning from large amounts of real-world data without needing ground-truth source waveforms \cite{Wisdom2020}. Although MixIT has been successfully adapted to other SE tasks, it requires access to the in-domain noise. To address this issue, Tzinis \textit{et al.} proposed RemixIT \cite{Tzinis2021,Tzinis2022}, which adopts a teacher-student training framework to achieve state-of-the-art (SoTA) performance on various unsupervised and semi-supervised SE tasks. The framework's flexibility allows using any SE model as the teacher model.

It is known that when the training data used for an SE model matches the test data, the performance of the test is higher, and vice versa. This makes domain matching a critical performance factor. This is especially important in real-world scenarios where the noise is complex, and it is challenging to synthesize similar noise during model training. We believe that this issue needs to be addressed urgently. Therefore, inspired by CTT, NyTT, and RemixIT, we propose a new training and inference strategy based on a teacher-student structure, which uses noisy and enhanced speech as a target for SE without any clean speech. The novelty of this study spans the following aspects:

\begin{figure*}[t]
\centering
\includegraphics[width=0.96\textwidth]{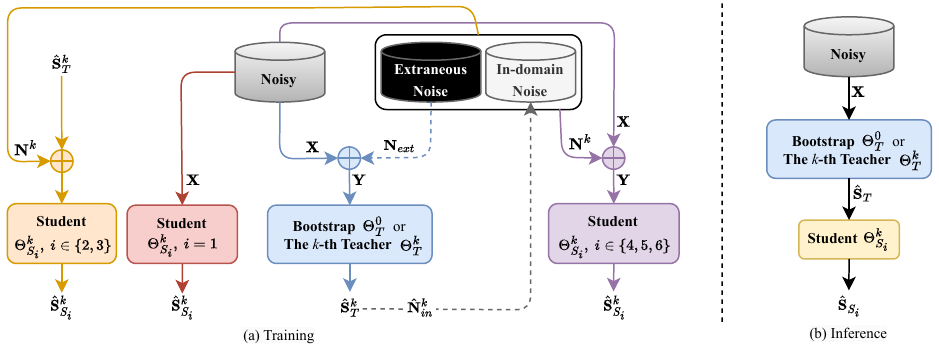}
\vspace{-5pt}
\caption{An overview of the proposed strategy for training and inference, where $\mathbf{X}$, $\mathbf{N}$, $\mathbf{Y}$, and $\mathbf{S}$ denote noisy speech, noise, synthesized noisy speech, and enhanced speech, respectively, in the sense of mini-batches. The blue dashed line only exists when Bootstrap $\Theta_T^{0}$ is trained by NyTT. The gray dashed line shows the flow of how the estimated in-domain noise $\hat{\mathbf{N}}_{in}$ is obtained from the $k$-th inference-in-training by Teacher $\Theta_T^{k}$, i.e., $\hat{\mathbf{N}}^k_{in}=
\mathbf{X}-\hat{\mathbf{S}}^k_T$. Students $\Theta^k_{S_i}\ (i=1,...,6)$ are elaborated in Section \ref{sec:training_framework}.}
\label{fig:structure}
\vspace{-15pt}
\end{figure*}

\begin{enumerate}[label=\arabic*),leftmargin=14pt]
\item Our strategy, called Ny/EnhTT\footnote{The code is open-sourced on \url{https://github.com/Sinica-SLAM/Ny-EnhTT}.}, can skillfully extract out the in-domain noise related components from noisy speech with the assistance of extraneous noise. This helps alleviate the effect of domain mismatch.
\item We explore several student models that vary concerning the use of enhanced speech, noisy speech, and the estimated in-domain noise. This helps the teacher model to extract more reliable in-domain noise. 
\item We discover that the resulting teacher-student model is more suitable for a strategy of inference called the teacher/student inference, where predicted clean speech is derived successively through the teacher and final student models. This makes the performance of NyTT further improved.
\end{enumerate}

\section{Related Work}
\label{sec:method}

\subsection{Noisy-target Training (NyTT)}
\label{ssec:nytt}

Traditionally, supervised SE methods are based on CTT, use clean speech as the training target, and the noisy input speech is synthesized by mixing clean speech and noise. Unlike CTT, in NyTT \cite{Fujimura2021a}, clean speech is replaced by noisy speech. As shown in Fig. \ref{fig:structure} (the blue part), the mini-batch $\mathbf{Y}\in \mathbb{R}^{B \times M}$ of the noisy input speech is synthesized by mixing the noisy speech $\mathbf{X} \in \mathbb{R}^{B \times M}$ and the noise $\mathbf{N} \in \mathbb{R}^{B \times M}$, where $B$ is the batch size, and $M$ is the signal length. The input $\mathbf{Y}$ is then fed into the model to get the estimated speech $ \hat{\mathbf{S}} \in \mathbb{R}^{B \times M}$. NyTT uses the mean squared error as the loss function to update the model in each iteration:
\begin{equation}
\label{eq:nytt_loss}
\mathcal{L}_{NyTT}=\frac{1}{B}||\mathbf{X}-\hat{\mathbf{S}}||_F^2,
\end{equation}
where $||\cdot||_F$ is the Frobenius norm.

\subsection{RemixIT}
\label{ssec:remixit}
RemixIT is a teacher-student training framework with SoTA results on several unsupervised and semi-supervised denoising tasks by adapting a model's domain to another domain \cite{Tzinis2022}. It uses the speech and noise the teacher model estimates to form paired training data for student model training. Specifically, the model trained with OOD data is the initial teacher model $\Theta_T^{0}$. When training $k$-th epoch, the teacher model $\Theta_T^{k}$ estimates the speech $\hat{\mathbf{S}}_T^k \in \mathbb{R}^{B \times M}$ and noise $\hat{\mathbf{N}}_{in}^k \in \mathbb{R}^{B \times M}$ from the in-domain noisy speech $\mathbf{X} \in \mathbb{R}^{B \times M}$:
\begin{equation}\label{eq:teacher_output}
(\hat{\mathbf{S}}_T^k, \hat{\mathbf{N}}_{in}^k) = \Theta_T^{k}(\mathbf{X}).
\end{equation}
Then, the new noisy speech $\mathbf{Y} \in \mathbb{R}^{B \times M}$ for training the student model is synthesized by mixing the estimated speech $\hat{\mathbf{S}}_T^k$, and the shuffled estimated noise $\tilde{\mathbf{N}}_{in}^k = \mathbf{P}\hat{\mathbf{N}}_{in}^k \in \mathbb{R}^{B \times M}$. $\mathbf{P}$ is a $B \times B$ permutation matrix used to generate random-order estimated in-domain noise from $\hat{\mathbf{N}}_{in}^k$.

The teacher model $\Theta_T^{k}$ is updated according to one of the following Teacher Update Protocols (TUPs):
\begin{itemize}[leftmargin=*,itemsep=0pt]
\item{\bf Static teacher}: The teacher model is not updated during the training of the student model.
\item{\bf Exponentially moving average teacher}: For each epoch, the teacher model is replaced by the weighted sum of the latest student model and the current teacher model, i.e., $\Theta_T^{k+1}=\gamma \Theta_S^{k}+(1-\gamma)\Theta_T^{k}$, where $\gamma = 0.005$.
\item{\bf Sequentially updated}: The teacher model is replaced by the latest student model every $K$ epochs. However, we do not consider this protocol our TUP because the results are worse than the baseline in the task of VoiceBank-DEMAND.
\end{itemize}

\section{Proposed Method}
\label{sec:ourmethod}

\subsection{Training Framework}
\label{sec:training_framework}

We use NyTT to train the initial teacher model. The blue lines in Fig. \ref{fig:structure} illustrate the training process of Bootstrap $\Theta_T^{0}$ and the inference-in-training process of the teacher model $\Theta_T^{k}$. The noisy speech $\mathbf{X}$ and the extraneous noise $\mathbf{N}_{ext} \in \mathbb{R}^{B \times M}$ are mixed into the noisy input speech $\mathbf{Y}$. Different from the loss function in Eq. \ref{eq:nytt_loss}, referring to the DEMUCS architecture \cite{Defossez2019}, the model is updated by minimizing the mean absolute error of the noisy speech $\mathbf{X}$ and the estimated noisy speech $\hat{\mathbf{S}}_T^k$.

Given the initial teacher model $\Theta_T^{0}$, the epoch size $E$, the number of mini-batch $N_m$, the batch size $B$, and TUP, we follow the teacher-student training process of RemixIT. In $k$-th epoch, we first sample a batch of noisy speech $X$ and initialize a random $B \times B$ permutation matrix for shuffling the estimated in-domain noise $\hat{\mathbf{N}}_{in}^k$. We then estimate speech and in-domain noise from $\mathbf{X}$ and remix in-domain noisy speech for student model training. After the student model is trained, the teacher model is updated according to the given TUP.

We propose six methods to train the student model, each of which, Ny/EnhTT-$i$, is for Student $\Theta_{S_i}^{k}$ in Fig. \ref{fig:structure}. Note that $\mathbf{N}^{k}$ in Fig. \ref{fig:structure} can be either samples from estimated in-domain noise, samples from estimated in-domain and extraneous noise, or mixtures of estimated in-domain and extraneous noise. The six methods are described as follows:
\begin{itemize}[leftmargin=*,itemsep=0pt]
\item{\bf Ny/EnhTT-1} (for Student $\Theta_{S_1}^{k}$) takes the noisy speech $\mathbf{X}$ as input (i.e., $\mathbf{Y}=\mathbf{X}$) and the speech, $\hat{\mathbf{S}}_T^k$, estimated by the teacher model as the target.
\item{\bf Ny/EnhTT-2} (for Student $\Theta_{S_2}^{k}$) takes the remix of estimated speech $\hat{\mathbf{S}}_T^k$ and noise $\mathbf{N}^{k}$ as input (i.e., $\mathbf{Y}=\hat{\mathbf{S}}_T^k+\mathbf{N}^{k}$) and $\hat{\mathbf{S}}_T^k$ as the target. $\mathbf{N}^{k}$ contains only estimated in-domain noise (i.e., $\mathbf{N}^{k}=\hat{\mathbf{N}}_{in}^k$).
\item{\bf Ny/EnhTT-3} (for Student $\Theta_{S_3}^{k}$) takes the remix of estimated speech $\hat{\mathbf{S}}_T^k$ and noise $\mathbf{N}^{k}$ as input (i.e., $\mathbf{Y}=\hat{\mathbf{S}}_T^k+\mathbf{N}^{k}$) and $\hat{\mathbf{S}}_T^k$ as the target. $\mathbf{N}^{k}$ is a mixture of estimated in-domain and extraneous noise (i.e., $\mathbf{N}^{k}=\hat{\mathbf{N}}_{in}^k + \mathbf{N}_{ext}$).
\item{\bf Ny/EnhTT-4} (for Student $\Theta_{S_4}^{k}$) takes the remix of noisy speech $\mathbf{X}$ and noise $\mathbf{N}^{k}$ as input (i.e., $\mathbf{Y}=\mathbf{X}+\mathbf{N}^{k}$) and $\mathbf{X}$ as the target. $\mathbf{N}^{k}$ contains only estimated in-domain noise (i.e., $\mathbf{N}^{k}=\hat{\mathbf{N}}_{in}^k$). Its complete training process is summarized in Algorithm \ref{alg:training}.
\item{\bf Ny/EnhTT-5} (for Student $\Theta_{S_5}^{k}$) takes the remix of noisy speech $\mathbf{X}$ and noise $\mathbf{N}^{k}$ as input (i.e., $\mathbf{Y}=\mathbf{X}+\mathbf{N}^{k}$) and $\mathbf{X}$ as the target.
Each sample in $\mathbf{N}^{k}$ is from $\hat{\mathbf{N}}_{in}^k$ and $\mathbf{N}_{ext}$.
\item{\bf Ny/EnhTT-6} (for Student $\Theta_{S_6}^{k}$) takes the remix of noisy speech $\mathbf{X}$ and noise $\mathbf{N}^{k}$ as input (i.e., $\mathbf{Y}=\mathbf{X}+\mathbf{N}^{k}$) and $\mathbf{X}$ as the target. $\mathbf{N}^{k}$ is a mixture of estimated in-domain and extraneous noise (i.e., $\mathbf{N}^{k}=\hat{\mathbf{N}}_{in}^k + \mathbf{N}_{ext}$).
\end{itemize}
Note that $\hat{\mathbf{N}}_{in}^k$ is ``predicted'' or ``estimated'' by the $k$-th teacher model, not ``real'' in-domain noise.

\subsection{Teacher/Student Inference}
\label{ssec:two-stageinference}

As mentioned in Section \ref{sec:intro}, the enhanced speech $\hat{\mathbf{S}}_T^0$ still contains some in-domain noise that cannot be removed entirely by the initial NyTT model $\Theta_T^{0}$. Although Grzywalski \textit{et al.} claim that performing speech enhancement through the same network up to five times improves speech intelligibility \cite{Grzywalski2022}, there is no significant improvement when we pass noisy speech through $\Theta_T^{0}$ twice (see Table \ref{tab:se_static_results}).

Instead of using \textit{the same} model in multi-stage inference, we successively use the initial teacher model $\Theta_T^{0}$ and the final model (i.e., $\Theta_{S_i}^{k}$ or $\gamma\Theta_{S_i}^{k}+(1-\gamma)\Theta_T^{k}$) for the teacher/student inference. The teacher model $\Theta_T^{0}$ enhances noisy speech in the first inference. Then, in the second inference, we feed the enhanced speech into $k$-th student model $\Theta_{S_i}^{k}$ or the model whose parameters are the weighted sum of the parameters of the kth teacher model and the kth student model, i.e., $\gamma\Theta_{S_i}^{k}+(1-\gamma)\Theta_T^{k}$. We were surprised to see a considerable improvement in this practice. We will show the effect of different $k$ values on the experimental results in Section \ref{ssec:results}.

\section{Experiments}
\label{sec:experiments}

\subsection{Datasets}
\label{ssec:dataset}

We used VoiceBank-DEMAND as the in-domain noisy speech dataset \cite{Valentini-Botinhao2016}. The training set consists of 28 speakers (11,572 utterances) with four signal-to-noise ratios (SNR: 15, 10, 5, and 0 dB). The test set consists of two speakers (824 utterances) with four SNRs (17.5, 12.5, 7.5, and 2.5 dB). We also used the CHiME-3 backgrounds as the extraneous (OOD) noise set \cite{Barker2015}. DEMAND and CHiME-3 backgrounds are part of the training noise set in the original NyTT study \cite{Fujimura2021a}, but DEMAND and CHiME-3 backgrounds were constructed from different authors, environments, recording devices, noise sources, etc. Hence, we think they are not in the same domain. There might be some similarities between them at the signal level, but we do not know.

\subsection{Model Structure}
\label{ssec:modelstracture}

We used DEMUCS as the model architecture \cite{Defossez2019}. It was developed for real-time SE in the waveform domain and has been widely adopted in academia and industry. It consists of a U-net connected encoder and decoder, and the configurable parameters are the number of layers ($L$) and the number of initially hidden channels ($H$). We upsampled the input audio by the resampling factor $U$, fed it to the encoder, and downsampled the model's output by the sampling rate of the original input. The $i$-th layer of the encoder consists of a convolutional layer with a kernel size of $K$, a stride of $S$, and $2^{i-1}H$ output channels, followed by a ReLU activation and a $1 \times 1$ convolution with an output channel of $2^{i}H$, and a GLU activation that converts the number of channels to $2^{i-1}H$. A sequence model between the encoder and decoder is an LSTM network with two layers (each with $2^{L-1}H$ hidden units). We adopted a \textit{causal} version of DEMUCS; therefore, the LSTM was unidirectional. The $i$-th layer of the decoder takes $2^{L-i}H$ channels as input. It performs $1 \times 1$ convolution of $2^{L-i+1}H$ channels, a GLU activation function for $2^{L-i}H$ output channels, a transposed convolution of kernel size $8$, stride $4$, and $2^{L-i-1}H$ output channels, and a ReLU function in sequence. There is no ReLU function in the last output layer. The experimental parameters are $U=4$, $S=4$, $K=8$, $L=5$, and $H=48$.

\begin{figure}[t]
\vspace{-10pt}
\begin{algorithm}[H]
\caption{Proposed Training Process for Ny/EnhTT-4}
\small
\label{alg:training}
\begin{algorithmic}[1]
\State \textbf{Given} the initial teacher model $\Theta_T^{0}$, the epoch size $E$, the number of mini-batch $N_m$, and the batch size $B$
\For{\text{$k \in \{0,\ldots,E$\}}}
\For{each batch \text{$batch_j,\:j = 1,\ldots,N_m$}}
\State Sample noisy speech $\mathbf{X} \gets \{\mathbf{x}_{i}\}^{B}_{i=1}$
\State $\mathbf{P} \gets$ Initialize a random $B \times B$ permutation matrix
\State Estimate speech using teacher $\hat{\mathbf{S}}_{T}^{k} \gets \Theta_T^{k}(\mathbf{X})$
\State Estimate in-domain noise $\hat{\mathbf{N}}_{in}^k \gets \mathbf{X} - \hat{\mathbf{S}}_{T}^k$
\State Remix in-domain noisy speech $\mathbf{Y} \gets \mathbf{X} + \mathbf{P}\hat{\mathbf{N}}_{in}^k$
\State Update student model $\Theta_{S_i}^{k} \gets \text{NyTT}(\mathbf{Y})$
\EndFor
\If {$TUP$ is Static}
\State Teacher model remains the same $\Theta_T^{k+1} \gets \Theta_T^{k}$
\ElsIf {$TUP$ is Exponentially moving average}
\State Update teacher model $\Theta_T^{k+1} \gets \gamma \Theta_{S_i}^{k} + (1 - \gamma)\Theta_T^{k}$
\EndIf
\EndFor
\end{algorithmic}
\end{algorithm}
\vspace{-25pt}
\end{figure}

\subsection{Training Details}
\label{ssec:trainingdetail}

All our models were trained by the Adam optimizer with a step size of $3 \times 10^{-4}$, a momentum of $\beta_1= 0.9$, and a denominator momentum $\beta_2 = 0.999$. We used the Shift, Remix, and BandMask data augmentation methods proposed by Défossez \textit{et al.} \cite{Defossez2019}. Shift is to apply a random shift from $0$ to $n$ seconds. Remix shuffles the noises in a batch to form new noisy mixtures. BandMask is a band-stop filter with a stop band between f0 and f1, sampled to remove $20\%$ of frequencies in the Mel scale. All audio is sampled at 16 kHz. We randomly chose an SNR between $-5$ and $5$ dB when mixing two signals.

The NyTT baseline was trained for 500 epochs using VoiceBank-DEMAND (noisy speech) and CHiME-3 (extraneous noise). This baseline NyTT model was also used as the initial teacher model. Each student model with the static teacher as TUP was trained for 500 epochs. Each student model with the exponentially moving average teacher as TUP was trained for 35 epochs. Because the training data did not contain clean speech, and there was no validation set for selecting the best model, we constantly tested the model after the last epoch in the experiments.

\begin{table}[t]
\caption{Results of the baseline and our proposed models, which use ``static teacher'' as TUP, on VoiceBank-DEMAND.}
\vspace{-5pt}
\label{tab:se_static_results}
\small
\setlength{\tabcolsep}{8pt}
\centering
\begin{tabular}{lcccc}
\toprule
\multirow{2}{*}{\bf{Method}} & \multicolumn{2}{c}{\bf{PESQ}} & \multicolumn{2}{c}{\bf{STOI}} \\
\cmidrule(lr){2-3} \cmidrule(lr){4-5}
& \bf{S} & \bf{T/S} & \bf{S} & \bf{T/S} \\
\midrule
NyTT & 2.20 & 2.21 & 0.932 & 0.932 \\
\midrule
Ny/EnhTT-1 & 2.04 & 2.22 & 0.923 & 0.932 \\
Ny/EnhTT-2 & 2.07 & 2.22 & 0.928 & 0.932 \\
Ny/EnhTT-3 & 2.10 & 2.20 & 0.928 & 0.930 \\
Ny/EnhTT-4 & 2.04 & 2.26 & 0.927 & 0.933 \\
Ny/EnhTT-5 & 2.10 & 2.26 & 0.928 & 0.932 \\
Ny/EnhTT-6 & \bf{2.19} & \bf{2.28} & 0.930 & 0.931 \\
\bottomrule
\end{tabular}
\vspace{-5pt}
\end{table}

\begin{table}[t]
\caption{Results of the baseline and our proposed models, which use ``exponentially moving average teacher'' as TUP, on VoiceBank-DEMAND.}
\vspace{-5pt}
\label{tab:se_ema_results}
\small
\setlength{\tabcolsep}{8pt}
\centering
\begin{tabular}{lcccc}
\toprule
\multirow{2}{*}{\bf{Method}} & \multicolumn{2}{c}{\bf{PESQ}} & \multicolumn{2}{c}{\bf{STOI}} \\
\cmidrule(lr){2-3} \cmidrule(lr){4-5}
& \bf{S} & \bf{T/S} & \bf{S} & \bf{T/S} \\
\midrule
NyTT & 2.20 & 2.21 & 0.932 & 0.932 \\
\midrule
Ny/EnhTT-1* & 2.20 & 2.36 & 0.927 & 0.928 \\
Ny/EnhTT-2* & \bf 2.22 & \bf 2.37 & 0.926 & 0.927 \\
Ny/EnhTT-3* & 2.11 & 2.27 & 0.923 & 0.926 \\
Ny/EnhTT-4* & \bf 2.22 & \bf 2.37 & 0.927 & 0.928 \\
Ny/EnhTT-5* & 2.15 & 2.31 & 0.924 & 0.927 \\
Ny/EnhTT-6* & 2.11 & 2.27 & 0.923 & 0.926 \\
\bottomrule
\end{tabular}
\vspace{-15pt}
\end{table}

\subsection{Results}
\label{ssec:results}

The results are shown in Table \ref{tab:se_static_results} and Table \ref{tab:se_ema_results}. The top row shows the results of the NyTT baseline, which is used as the initial teacher model for training our proposed models. Table \ref{tab:se_static_results} shows the results of student models trained with a static teacher. Table \ref{tab:se_ema_results} shows the results of student models trained with an exponentially moving average teacher. \textbf{S} refers to only using the student model for inference. \textbf{T/S} refers to using the initial teacher model for the first inference and the student model for the second inference. Two standardized metrics were used to evaluate the SE performance: perceptual evaluation of speech quality (PESQ) \cite{Rix2001}, and short-time objective intelligibility measure (STOI) \cite{Taal2011}.

Table \ref{tab:se_static_results} shows that all student models trained with the static teacher performed worse than the baseline in single-stage inference in terms of PESQ. However, almost all student models outperformed the baseline in the teacher/student inference. Notably, the best performer among these models is Ny/EnhTT-6, trained with a mixture of estimated in-domain and extraneous noise as input. The mixture of estimated in-domain and extraneous noise is relatively similar to the training noise of the baseline, thereby giving Ny/EnhTT-6 comparable performance to the baseline. Compared with other models, Ny/EnhTT-6 showed less improvement in the teacher/student inference. A similar trend was also observed in the results of Ny/EnhTT-3.

Comparing Table \ref{tab:se_static_results} and Table \ref{tab:se_ema_results}, we found the student models trained with the exponentially moving average teacher outperformed the student models trained with the static teacher in terms of PESQ. Some models outperformed the baseline in single-stage inference, and all models performed well in the teacher/student inference. However, in terms of STOI, the student models trained with the exponentially moving average teacher were worse than those trained with the static teacher and the baseline, which requires further study.

Table \ref{tab:se_results_with_others} compares our best model with previous supervised and unsupervised models. It can be seen that the supervised models are still more capable than the unsupervised models. When comparing the unsupervised models, our self-implemented NyTT model and our best Ny/EnhTT-4* model were inferior to the NyTT model in single-stage inference \cite{Fujimura2021a}. Possible reasons are as follows. First, the model structure is slightly different. Second, we used a causal model architecture, which is less effective than a non-causal one but more practical in real-world applications. Third, the original NyTT study used more training data than this work.

Moreover, we explore the effect of different teacher models $\Theta_T^k$ in the teacher/student inference. In Table \ref{tab:2-stage_exp}, we find that when $k \leq 2$, the overall performance slightly increases, but the results are worse afterward (when $k>3$). The closer $\Theta_T^k$ is to the final student model (when $k$ is larger), the worse the performance is.

\begin{table}[t]
\caption{PESQ achieved by various existing (un)supervised SE methods on VoiceBank-DEMAND. $\dagger$ indicates that external noisy data is used during training.}
\vspace{-6pt}
\label{tab:se_results_with_others}
\small
\centering
\scalebox{0.90}{
\begin{tabular}{lccc}
\toprule
\bf{Method} & \bf{Clean Speech?} & \bf{S} & \bf{T/S} \\
\midrule
SEGAN \cite{Pascual2017} & \ding{51} & 2.16 & N/A \\
MetricGAN \cite{Fu2019} & \ding{51} & 2.86 & N/A \\
DEMUS \cite{Defossez2020} & \ding{51} & 3.07 & N/A \\
CMGAN \cite{Cao2022} & \ding{51} & 3.41 & N/A \\
\midrule
NyTT$^{\dagger}$ \cite{Fujimura2021a} & \ding{55} & 2.30 & N/A \\
MetricGAN-U (full) \cite{Fu2022} & \ding{55} & 2.13 & N/A \\
NyTT (our implementation) & \ding{55} & 2.20 & 2.21 \\
\bf Proposed (Ny/EnhTT-4*) & \ding{55} & 2.22 & \bf 2.37 \\
\bottomrule
\end{tabular}}
\vspace{-5pt}
\end{table}

\begin{table}[t]
\caption{PESQ achieved by different teacher models $\Theta_T^k$ used the teacher/student inference on VoiceBank-DEMAND. The model of Ny/EnhTT-4 with ``exponential moving average teacher'' as TUP is the student model.}

\vspace{-6pt}
\label{tab:2-stage_exp}
\small
\setlength{\tabcolsep}{6pt}
\centering
\begin{tabular}{ccccccc}
\toprule
\bf k & 0 & 2 & 4 & 6 & 8 & 10\\
\midrule
\bf PESQ & 2.371 & \bf{2.372} & 2.368 & 2.366 & 2.363 & 2.359 \\
\bottomrule
\end{tabular}
\vspace{-15pt}
\end{table}

\section{Conclusions and Future Work}
\label{sec:conclusion}
In this paper, we have proposed a training and inference strategy for SE to mitigate the shortcomings of NyTT and other supervised methods. Our proposed method combines CTT, NyTT, and RemixIT and uses enhanced speech to estimate in-domain noise. Our experiments show that the exponentially moving average is the best teacher protocol for unsupervised SE tasks. It is also found that the teacher/student inference helps our proposed framework further to improve the performance in terms of PESQ and STOI. In the future, more powerful models, such as a non-causal DEMUCS or its transformer-based version, will be implemented in this framework. Furthermore, we will analytically investigate why the teacher/student inference leads to a performance boost. It can provide another perspective for designing more efficient teacher-student frameworks without the need for multiple stages of inference.

\bibliographystyle{IEEEtran}
\bibliography{references.bib}

\end{document}